Person to contact for this submission: M K Tiwari
(mktiwari@rrcat.gov.in ; phone: +91-731-2442124; fax: +91-731-2442140)

Article

# Temporal Coherence in Surface x-ray Standing Waves

M K Tiwari* and Gangadhar Das

*Indus Synchrotrons Utilization Division, Raja Ramanna Centre for Advanced Technology,*

*Indore-452013 (M P), India*

We present a unified elucidation for the role temporal coherence properties of x-rays in surface x-ray standing wave analysis, under total external reflection condition. Analytical formulas have been derived that provide a direct relation between the intensity distribution of x-ray standing wave (XSW) pattern and complex degree of coherence of the incoming x-rays. The novelty of the proposed method has been demonstrated experimentally together with the theoretical results for a real application by analyzing the distribution of metal nanoparticles on a Si mirror surface. Our results demonstrate that the modified XSW approach opens up an opportunity to unambiguously interpret structural studies of large dimension nanostructure materials of height 200nm (*or more*), deposited on a substrate surface.

*Corresponding Author

Over the past decades, XSW technique has been an exciting frontier that promises nondestructive evaluation of impurity atoms at nanometer to angstrom resolutions. The method has been widely exploited for a large number of applications including determination of positions of impurities in crystals[1-3], dispersion of absorbed atoms[4-8], distribution of metal nanoparticles on surfaces[9-11] as well as to study the interface structure and density variations in synthetic periodic multilayer structures [12,13].

However, despite these remarkable advancements in both fundamental understanding and practical applications of the technique, the influence of a few unknown is still widely lacking. A largely unexplored area in the XSW method is the realisation how temporal coherence properties of x-rays influence the XSW field intensity distribution above a reflector surface. In the conventional XSW approach[14-17], the structures of the complex molecules or layered materials are unraveled by considering a perfectly monochromatic x-ray source with zero wavelength dispersion ($\Delta\lambda \approx 0$) in the model calculations, thus assuming an infinite longitudinal coherence length of the incoming x-rays. However, in practice no x-ray source is completely monochromatic, therefore the above assumption remains no longer valid if the XSW measurements are accomplished using an x-ray source characterized by a specific distribution in intensity and a finite spectral bandwidth of the wavelengths. The understanding of consequences of source properties in XSW induced glancing incidence angle dependent fluorescence analysis is nontrivial for its potential utility in case of study of especially large dimension nanostructured materials and hence needs to be addressed properly. It is well established that the temporal coherence is strongly correlated with spectral bandwidth of the source. Thus, in order to fully describe the XSW field produced by an x-ray beam of finite spectral bandwidth, it is evidently desirable to consider its temporal coherence properties. Inspired by these gaps on theoretical explanation in the existing literature we step towards this goal.

Here, we detail how temporal coherence properties of x-rays influence the XSW field pattern under total external reflection condition on a mirror surface. We develop a detailed numerical description for the computation of the angle-dependent XSW field pattern by considering an x-ray beam characterized with a Gaussian distribution in intensity and a finite spectral bandwidth. The new approach allows us to perform XSW measurements with improved reliability using the currently available x-ray sources such as laboratory x-ray tubes and synchrotron radiation sources, which offer specific spectral resolutions for the incoming x-rays. To explicitly demonstrate the coherence aspects of the modified XSW approach, we perform an XSW characterization of metal nanoparticles dispersed on a polished Si mirror surface. Given that, the present approach includes source properties in the calculation of XSW field, it thus gives more accurate spatially depth resolved sensitivities with angstrom level resolution through structures of up to several hundreds of nanometers in height above a surface.

Figure 1 schematically describes the formation of an XSW field pattern in the overlap region of incident and reflected x-ray waves under the total external reflection condition on a mirror surface where we assume the electric field vectors $\mathbf{E}_0^i$ and $\mathbf{E}_0^r$ respectively represent the incident and reflected planar x-ray wave fields on a smooth reflecting surface. Then, the instantaneous field states for the two plane waves can be expressed by

$$\mathbf{E}^i(\mathbf{r},t) = \mathbf{E}_0^i \exp[i(\omega \times t - \mathbf{k}\cdot\mathbf{r})]$$

$$\mathbf{E}^r(\mathbf{r},t) = \mathbf{E}_0^r \exp[i(\omega \times t - \mathbf{k}'\cdot\mathbf{r})]$$

Where $\mathbf{k}$ and $\mathbf{k}'$ respectively represent the wave vectors for incident and reflected waves $\left(|\mathbf{k}| = |\mathbf{k}'| = 2\pi/\lambda\right)$, λ is the wavelength of the incident radiation. Here we assume that two plane waves are coherent and thus their traveling frequency, ω = 2πν , does not change during reflection and $\mathbf{r}$ is the position vector in the X-Z plane. It is convenient to express the net electromagnetic field intensity at height z above a reflecting surface as

$$I(\theta,z) = \left|\mathbf{E}_0^i + \mathbf{E}_0^r\right|^2 = I_0\{1 + R + 2\sqrt{R}\cdot \mathrm{Re}[\gamma(\Delta)]\} \tag{1}$$

Here $R = \left|\frac{E_0^r}{E_0^i}\right|^2$ represents the reflectivity above the substrate surface, $I_0 = \left|E_0^i\right|^2$ is the incident intensity, $\gamma(\Delta) = |\gamma(\Delta)|\cdot\exp(i[\upsilon(\theta) - H_z \times z])$ denotes the complex degree of temporal coherence[18]. Its value varies from 0 to 1. $H_z = \frac{4\pi}{\lambda}\sin(\theta)$ signifies is the real part of z-component of wave vector $\mathbf{H} = \mathbf{k}' - \mathbf{k}$, $\upsilon(\theta)$ is the phase of the E-field ratio and $\theta$ is the glancing angle of incidence on the substrate. The value of $\gamma$ strongly depends on the path difference $\Delta = AP - BP \approx 2z\sin(\theta)$ between incident and reflected x-ray waves at point P with a vertical height z above the substrate surface. It is expressed by

$$\gamma(\Delta) = \frac{\left\langle \mathbf{E}_0^i(P,t)\cdot\mathbf{E}_0^r(P,t)\right\rangle}{\left[\left\langle\left|\mathbf{E}_0^i(P,t)\right|^2\right\rangle\cdot\left\langle\left|\mathbf{E}_0^r(P,t)\right|^2\right\rangle\right]^{1/2}}$$

The mathematical details for the derivation of $\gamma$ for a quasi monochromatic incident x-ray beam characterized with a Gaussian spectral distribution are given elsewhere[19].

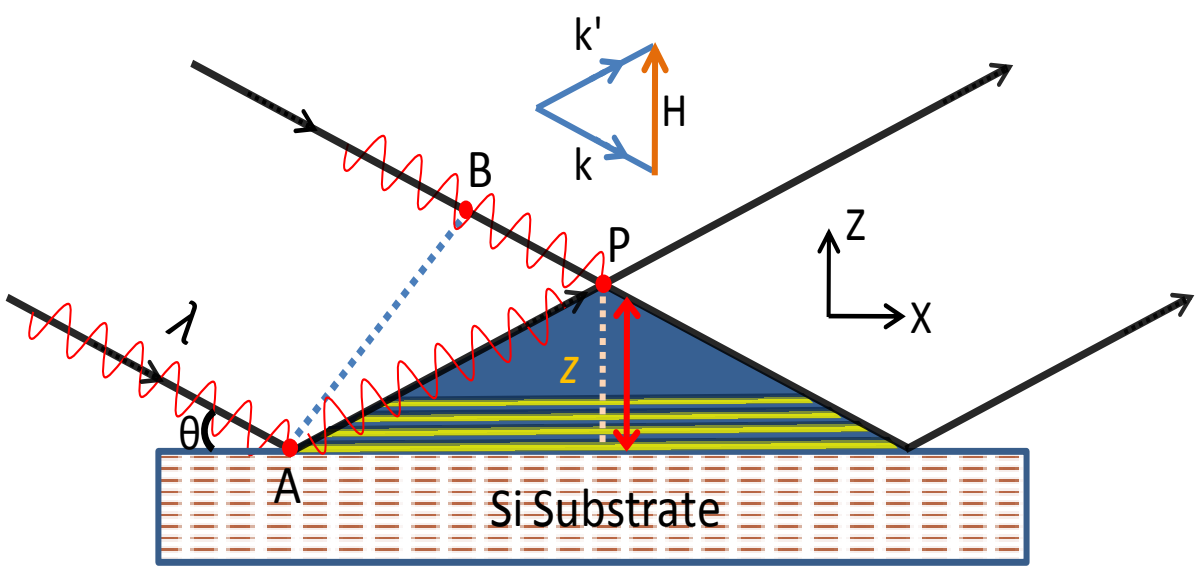

**FIG. 1.** (color online). A schematic illustration for the formation of XSW field under total external reflection of x-rays on a mirror surface.

Next, we perform numerical simulations to investigate the effect of temporal coherence in surface XSW measurements.

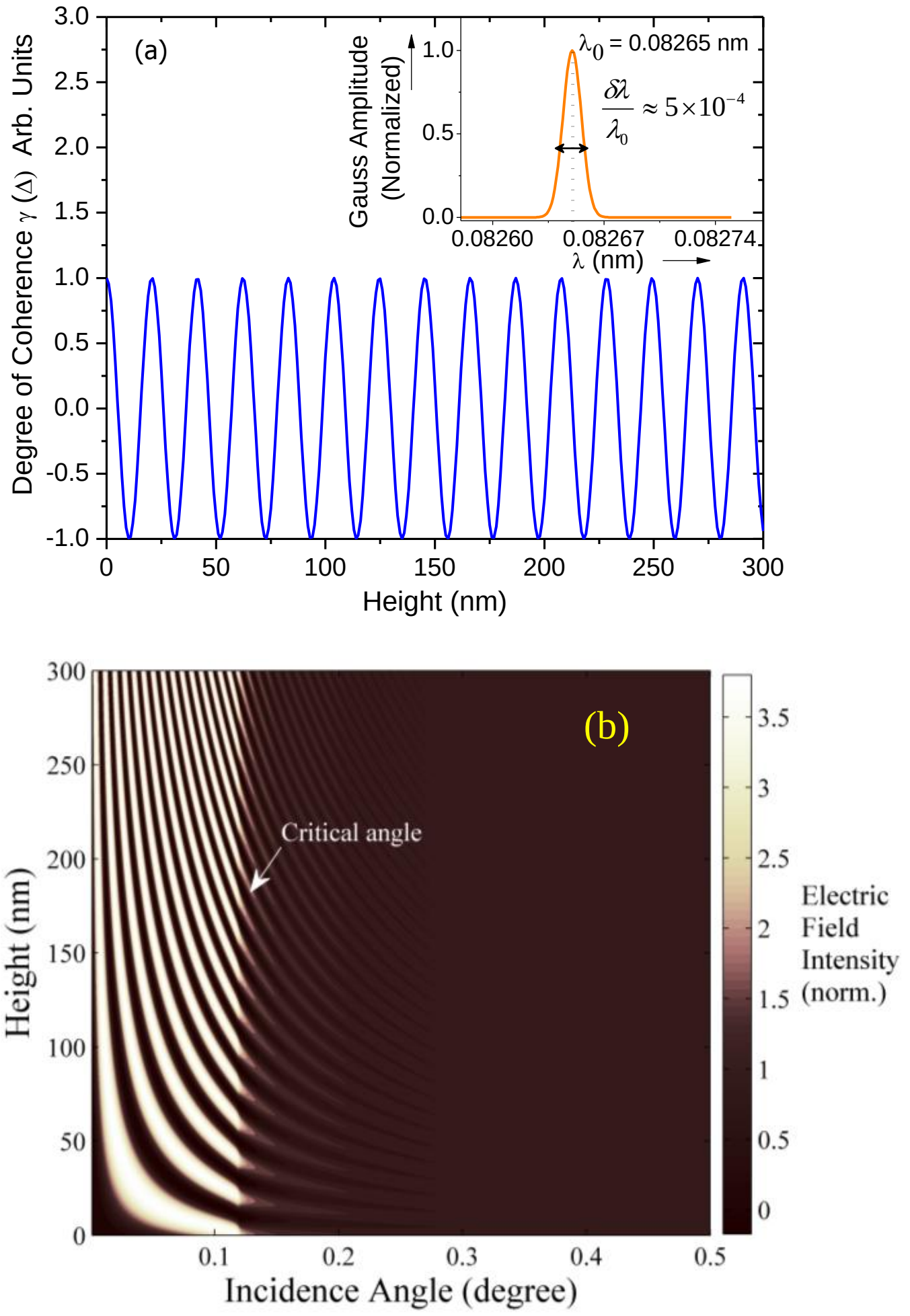

**FIG. 2.** (Color online) (a) Variation of the complex degree of coherence $\gamma(\Delta)$ as function of height z above the Si substrate surface. (b) Computed x-ray field intensity distribution above the Si substrate surface at an incident x-ray energy of 15 keV.

Figure 2a depicts co-sinusoidal variation of the complex degree of coherence function $\gamma(\Delta)$ as a function of the path length difference, represented in terms of height z above the substrate surface. In the simulation we have assumed a spectral resolution of the incoming x-ray beam $\delta\lambda/\lambda_0 \approx 5\times10^{-4}$ described with a Gaussian distribution profile as shown in the inset of the Fig.2a. At a low value of spectral resolution, the temporal coherence (the amplitude of $\gamma$) of the x-ray beam is preserved up to the height z = 300 nm above the substrate surface. Here we have assumed incident excitation energy of 15 keV and a fixed incidence angle $\theta \approx 0.114^0$ for the incoming x-rays. Figure 2b shows distribution of the total x-ray field intensity as a function of height z and incidence angle $\theta$ above the Si substrate surface. In the contour plot of the x-ray field intensity distribution (Fig. 2b), one clearly observes a boundary for the critical angle of the Si substrate. The visibility contrast of x-ray field intensity is significantly higher below the critical angle and exists in the form of interference fringes.

To precede further inline we extend our discussion to explain how relatively a large spectral width of the incoming x-ray beam influences the XSW field pattern in the total external reflection region. Figure 3a reports simulated profile of an exponentially decreasing oscillatory behavior of $\gamma(\Delta)$ with respect to the height z above the substrate surface, assuming a spectral resolution of $\delta\lambda/\lambda_0 \approx 8\times10^{-2}$. For large path difference values, for example, at z = 150 nm, incident and reflected beams hardly remain coherent with each other. As a result the interference term in Eq. (1) vanishes and the total x-ray intensity above the substrate surface simply represents the sum of the intensity of two beams. Figure 3(b) illustrates contour plot for the x-ray field intensity distribution above the substrate surface for an x-ray beam comprised of a spectral resolution $\delta\lambda/\lambda_0 \approx 8\times10^{-2}$. It can be seen here that the XSW modulations below the critical angle are strongly correlated with the complex degree of temporal coherence $\gamma(\Delta)$. If the path length difference $(\Delta)$ between two interfering beams (in terms of height z), is significantly smaller than the coherence length $L_T = \frac{2\log_e(2)}{\pi}\frac{\lambda}{(\delta\lambda/\lambda_0)}$ of the incident x-ray beam (i.e. $\Delta \ll L_T$), then a bright XSW fringe pattern is produced in the total reflection region. As the path length difference between the incident and reflected beams is increased, the visibility contrast of the XSW fringes is reduced. For example, for $\delta\lambda/\lambda_0 \approx 8\times10^{-2}$ and $L_T$ = 0.5 nm, corresponds to a path length difference value of z ≈ 75 nm, and the visibility of the XSW fringes reduced to ~ 50%. In situations where the path length difference values are significantly large (z >> $L_T$), the interference effect completely vanishes and the net intensity below the critical angle becomes equal to the sum of the intensities of the two beams.

The concepts of temporal coherence in the optical region are well understood. However, the effect of spectral resolution of the incoming x-rays in the computation of interference pattern and visibility of the XSW field on the top of a mirror surface are implemented by us for the first time. Nevertheless, it is very helpful and permits one to calculate XSW induced fluorescence profiles for any kind of nanostructure materials or surface layers of heights greater than 200 nm, which are directly deposited on a substrate surface. Below, we briefly address one such practical application of our newly developed technique to analyze the average vertical height of metal nanoparticles dispersed on a mirror surface.

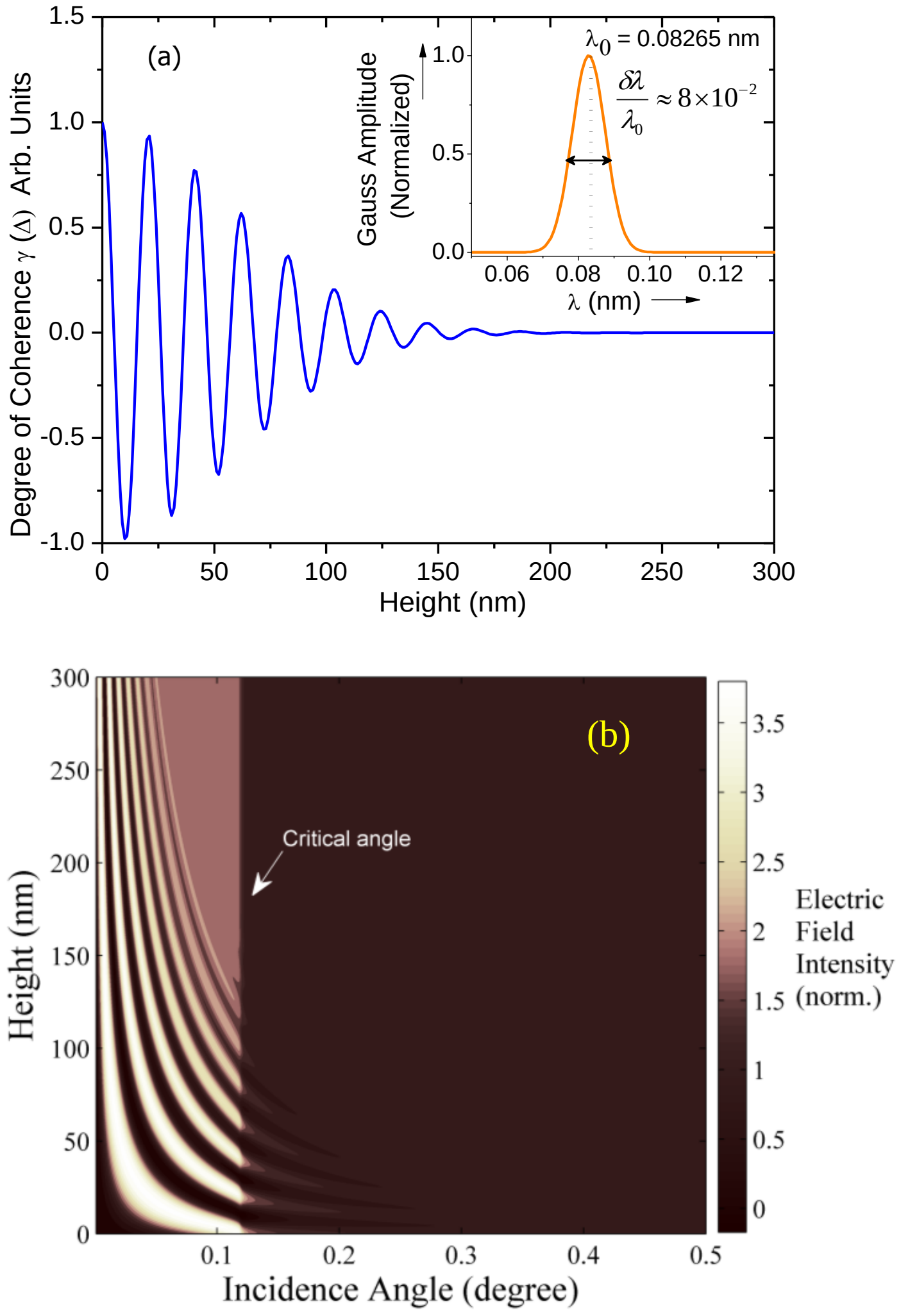

**FIG. 3.** (Color online) (a) Calculated complex degree of coherence $\gamma(\Delta)$ as function of height z above the Si substrate surface for an x-ray beam comprised of spectral resolution $\delta\lambda/\lambda_0 \approx 8\times10^{-2}$. (b) Computed x-ray field intensity distribution above the surface at x-ray energy of 15 keV.

*Nanoparticle distribution* − Figure 4 shows the results of simulations, demonstrating deeper insights on how monochromaticity of the incoming x-ray beam affects the sensitivity of the XSW enhanced angle dependent fluorescence profile in the case of a distribution of Au nanoparticles on a Si surface. The XSW enhanced angle-dependent fluorescence profiles are computed for the spherical shaped Au nanoparticles of diameter ≈ 250 nm, considering different spectral widths of the incident x-ray beam. We note here that the XSW fluorescence profile of the nanoparticles undergoes slowly with a distinct and systematic variation as we allow the monochromaticity of the incident x-ray beam to vary in the range 0 to $8\times10^{-2}$. The peak height of the XSW induced oscillations in the fluorescence profile decreases gradually below the critical angle $(\theta < 0.114^0)$, as we increase the spectral width of the incident x-ray beam. At spectral resolution of $\delta\lambda/\lambda_0 \approx 8\times10^{-2}$, even many XSW oscillations vanish from the fluorescence profile.

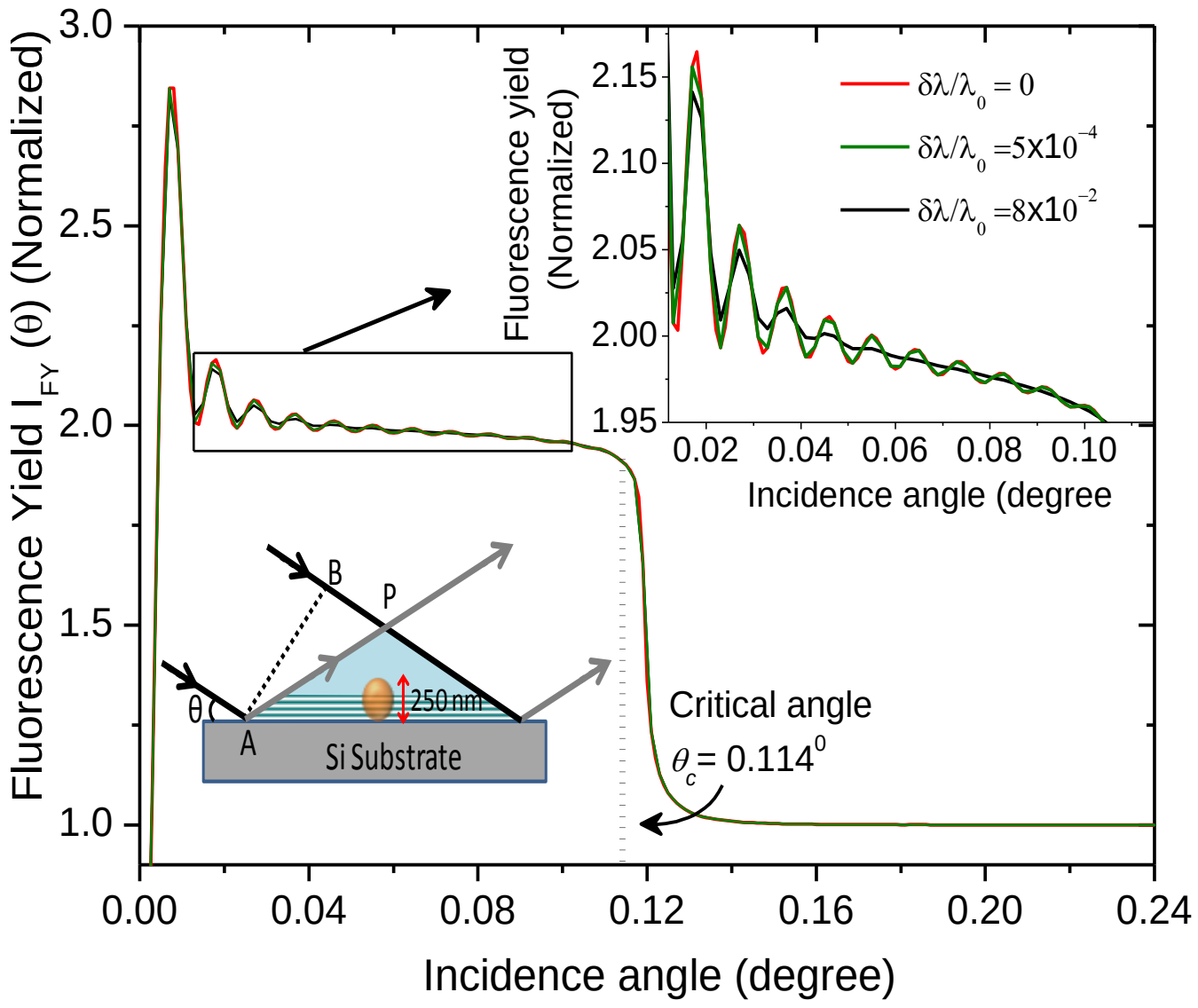

**FIG. 4.** (Color online) Computed XSW induced fluorescence profiles for spherical Au nanoparticles distribution on a Si substrate surface at different spectral resolution widths of the incoming x-ray beam. The angle dependent fluorescence profiles are simulated using Eq. (1) and the formula $I_{FY}(\theta)=\int_0^D I(\theta,z)\times\exp\left(-\frac{\mu_{Au}}{\sin(\varphi)}z\right)\times F(z)dz$, where $\mu_{Au}$ describes the linear absorption coefficient of Au for the outgoing fluorescent x-ray photons, φ is the take off angle of emitted fluorescent photons and D is the diameter of the nanoparticle. $F(z)=4z\left(\frac{D-z}{D^2}\right)$ represents the form factor for a spherical shape nanoparticle as a function of height z above the substrate surface[11].

In the inset of Fig.4, we have shown an expanded view for the simulated results below the critical angle of the substrate which clearly demonstrates the effect of spectral resolution on the visibility contrast of XSW induced fluorescence oscillations. These results are consistent and provide a pertinent theoretical explanation for the experimental observations reported by M J Bedzyk[8].

The experimental measurements for Au nanoparticles comprising of an average particle size of ~ 90 nm (dia.), dispersed on a Si(100) substrate surface were carried out at the B16 Test beamline of Diamond Light Source[20], UK. Incident x-rays of energy 15 keV, monochromatized with a fixed exit Si (111) double-crystal monochromator was used for fluorescence excitation of nanoparticle samples at grazing incidence angles. The fluorescent x-rays emitted from the Au nanoparticles were collected by a vortex spectroscopy detector placed normal to the sample surface through an Aluminum collimator (dia. ≈ 2 mm). Figure 5a depicts experimentally measured XSW induced Au-Lα fluorescence profile of the Gold nanoparticles along with the fitted profile as shown by the solid red line. For fitting of experimental results we considered the measured spectral resolution $(\delta\lambda/\lambda_0 \approx 1.4\times10^{-4})$ at 15 keV x-ray energy.

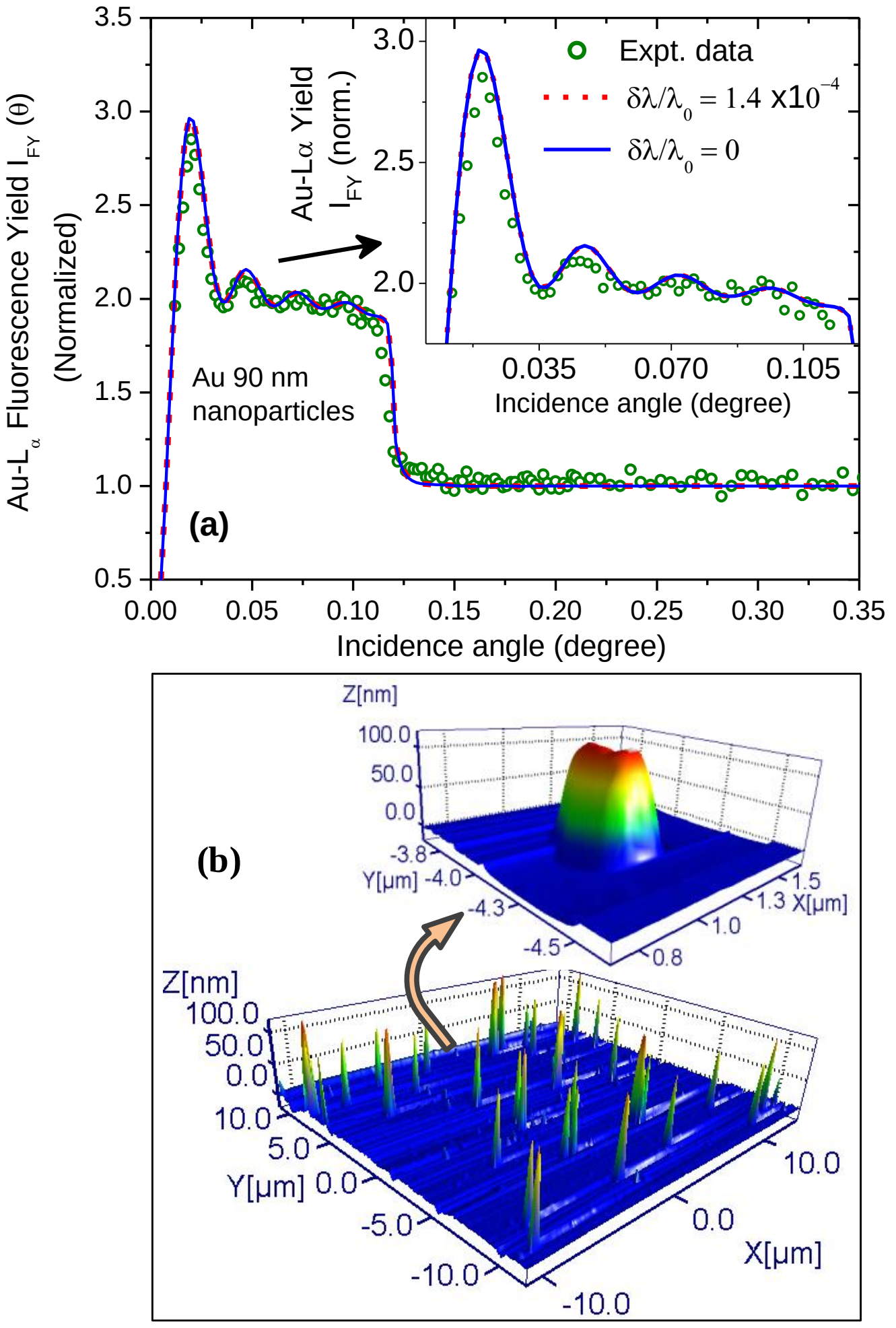

**FIG. 5.** (Color online) (a) Measured and fitted angle-dependent x-ray fluorescence profiles of Au nanoparticles, dispersed on top of a Si substrate at an x-ray energy of 15.0 keV. The curves of Au-Lα fluorescence yield have been normalized at an incidence angle of 0.3°. (b) AFM image of the distribution of Au nanoparticles on the Si substrate. The inset shows an expanded image of one nanoparticle.

Figure 5(a) shows a comparison of the measured and fitted Au-Lα fluorescence profiles. One observes several interference oscillations in the measured XSW induced Au-Lα fluorescence profile below the critical angle of the Si substrate $(\theta_c \approx 0.114^0)$. The peaking interference behavior in the fluorescence profile emerges due to the coincidence of the XSW antinodes within the volume of the Gold nanoparticles. The best fitted XSW fluorescence profile matching to the experimental data gives an average vertical height of the Au nanoparticles ~ 90 ± 1.0 nm, for their distribution over the Si mirror surface. In Fig.5a, we have also compared measured Au-Lα fluorescence profile along with the profiles computed assuming a perfectly monochromatic x-ray beam ($\delta\lambda/\lambda_0 = 0$), shown by the solid blue line. The inset of Fig. 5a shows an expanded view in the total external reflection region. It can be seen from Fig.5a that the computed fluorescence profile at spectral resolution of $\delta\lambda/\lambda_0 = 1.4\times10^{-4}$ closely agrees with the profile that obtained assuming a perfectly monochromatic x-ray source (i.e. $\delta\lambda/\lambda_0 = 0$). This due to the fact that at higher spectral resolutions, the incident and reflected x-ray beams remain fully coherent up to a height of several hundred of nanometers above the substrate surface. If the size of a nanostructure is considerably smaller than the coherence length (i.e. $D_{nanostructure} << L_T$), for example, in our case ~ 90nm nanoparticles, then full volume of a nanostructure is excited by the XSW field created with a 100% coherent superposition of the incident and reflected x-ray beams. The measured fluorescence profile deviates from the ideal behavior ($\delta\lambda/\lambda_0 = 0$) in situations if spectral resolution of the incident x-rays deteriorates or else if the height of a nanostructure increases on top of the substrate surface (see Fig. 4). All these factors increase an inconsistency of mutual coherence between incident and reflected x-ray beams.

It might be interesting here to also comment on the role of ‘coherent fraction’, an analogous ability that used to describe the distribution of adatoms on a surface in the crystal assisted XSW technique. The ‘coherent fraction’ ($f_H$) is essentially a system (or a specimen) specific quantity, which describes the coherent ordering of the impurity adatoms or the electron density distribution of molecules relative to the XSW field pattern. A perfectly ordered adatoms layer ($f_H$ =1) produces normalized fluorescence yield close to 4, whereas a random distribution of ad-atoms ($f_H$ =0) gives normalized fluorescence yield ≈ 2 in the XSW assisted fluorescence measurements. On the contrary, the temporal coherence of the incident x-ray beam dictates the coherent ordering of the XSW field pattern itself. The poor monochromaticity of the x-ray beam considerably affects the XSW field intensity distribution on a mirror surface as has been described by us previously.

In order to confirm the surface topography of the Au nanoparticles distribution on the Si surface, atomic force microscopy (AFM) measurements were carried out. Figure 5(b) shows a measured AFM image for the Au nanoparticles, which clearly demonstrates a mono-dispersed distribution of the nanoparticles. Inset of Fig.5b shows an expanded view for a single nanoparticle. AFM measurements give the average vertical size of the Au particles to be ~ 91 ± 1.0 nm. We find an encouraging level of agreement between the AFM and XSW measurements. To this end, we have investigated our methodology to distinguish the distribution of relatively small size Au nanoparticles (dia.≈ 90nm) over a Si substrate surface. However, it may be noted that the present approach can consistently be used to study nanostructures starting from the angstrom length scale and extending to hundreds of nanometers in heights.

To conclude, we have proposed, implemented and analyzed the temporal coherence properties of x-rays in the XSW measurements under total external reflection conditions on a mirror surface. Analytical formulas have been derived that take into account the source properties such as spectral resolution and its intensity distribution in the computation of fluorescence-induced XSW profiles. In contrast to the conventional approach where the XSW analysis is performed assuming a perfectly monochromatic x-ray beam, the new approach offers two evident advantages. First, it provides more sensible way of model calculations to fit the experimental data, thus leading us to derive structural parameters of a nanostructured material with improved accuracies. Second, our method proposes a generalized approach therefore it opens an opportunity to perform XSW measurements using any kind of monochromatic x-ray beams- be it produced from a crystal or a multilayer monochromator using synchrotron or laboratory x-ray sources. Finally, we emphasize that the methodology presented here seems strong enough to be a major step forward towards the study of large dimensional nanostructured materials in surface condensed matter physics applications, in particular examples of; Langmuir-Blodgett films, metal nanoparticles, structured surface layers, and periodic multilayers deposit on surfaces.

The authors would like to thank Prof. Donald Bilderback, Cornell High Energy Synchrotron Source, USA, for fruitful discussions and suggestions. Thanks also go to Dr. K J S Sawhney, Diamond Light Source (DLS) UK for providing successful beamtime and uninterrupted support. S. G. Alcock, DLS is gratefully acknowledged for the atomic force microscopy measurements on the Au nanoparticle samples.

## Outlook

We have shown that temporal coherence has an ability to efficiently modulate the intensity of the XSW field pattern, produced under the total reflection condition on a mirror surface. The XSW analysis of the nanostructured materials therefore may yields erogenous interpretations if the temporal coherence properties of the primary x-ray beam are not taken into account in the model calculations. Although, the x-ray standing wave XSW technique (*a non destructive probe for surface analysis*) has witnessed considerable progress and been exploited previously for a large number of applications in surface condensed matter physics, however the consequences of temporal coherence in XSW analysis were not fully explored. With the advent of many state of the arts synchrotron and x-ray free electron laser facilities, providing unprecedented level of temporal and spatial coherence properties of x-rays, such issues are of critical importance.

Present work is the first attempt to map the coherence properties in XSW analysis and provide a detailed mathematical description for the computation of angle dependent XSW fluorescence profile for nanostructured materials by taking into account source properties of the incoming x-rays (such as *spectral width* and *intensity distribution*), which otherwise are inaccessible from the existing literature. We clearly demonstrate the accuracy and novelty of the proposed approach through simulations and by experimental results. A further point to note here is that the discussion presented in our article is general and does not restrict to a specific x-ray source, therefore opens up an opportunity to perform XSW analysis using laboratory sources as well (*providing medium energy resolution and high photon flux for the incoming x-rays while synthetic multilayer's are used as monochromators*). Our work presents a new strong field regime for the characterization of nanoscaled systems. The authors believe that many researchers and practitioners of the x-ray standing wave field could be benefitted and would provide a step forward towards the accurate structural analysis of relatively large dimension nano-scaled systems using XSW method.

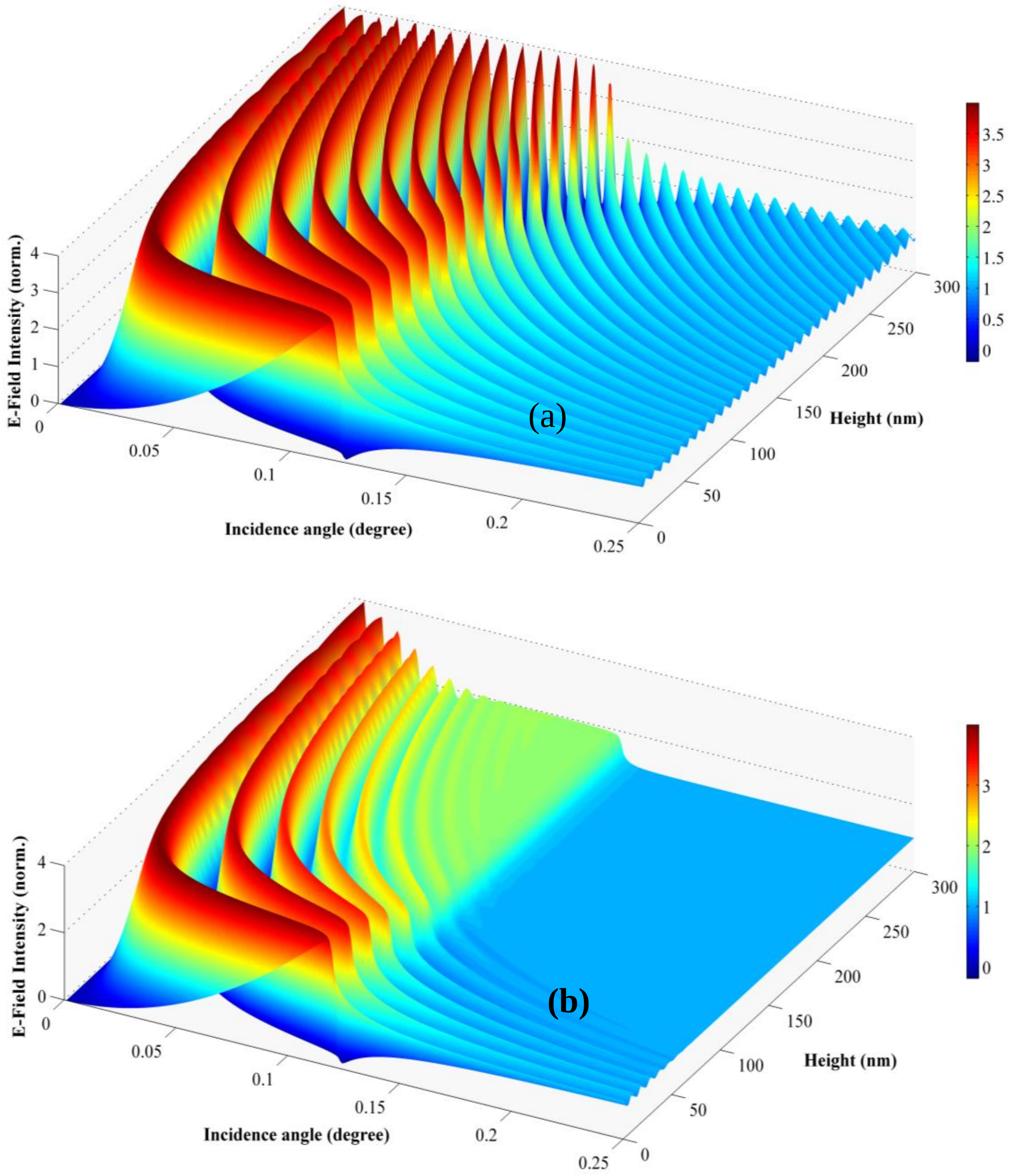

**FIG .** 3D view of the computed electric field intensity distribution above a Si mirror surface at different spectral resolutions of the incoming x-ray beam (*as shown in Figs.2b and 3b*) (a) $\delta\lambda / \lambda_0 = 1.4\times10^{-4}$, and (b) $\delta\lambda / \lambda_0 = 8\times10^{-2}$.

**Article Status**: *to be published shortly*